# Dimer Crystallization Induced by Elemental Substitution in the Honeycomb Lattice of Ru$_{1-x}$Os$_x$Cl$_3$

Kota Kataoka[1], Dirk Wulferding[2,3], Takeshi Yajima[1], Daisuke Nishio-Hamane[1], Daigorou Hirai[1], Seungyeol Lee[4], Kwang-Yong Choi[5], and Zenji Hiroi[1]

[1]*Institute for Solid State Physics, University of Tokyo, Kashiwa, Chiba 277-8581, Japan*
[2]*Center for Correlated Electron Systems, Institute for Basic Science, Seoul 08826, Republic of Korea*
[3]*Department of Physics and Astronomy, Seoul National University, Seoul 08826, Republic of Korea*
[4]*Department of Physics, Chung-Ang University, Seoul 06974, Republic of Korea*
[5]*Department of Physics, Sungkyunkwan University, Suwon 16419, Republic of Korea*

Substitution effects of Os for Ru in α-RuCl$_3$ are investigated in a wide composition range of $0 \leq x \leq 0.67$ in Ru$_{1-x}$Os$_x$Cl$_3$ by X-ray and electron diffraction, magnetic susceptibility, heat capacity, and Raman spectroscopy measurements. Apart from the Kitaev physics with antiferromagnetic interactions increasing with $x$, a rich phase diagram is obtained, which includes an antiferromagnetic long-range order below 12 K for $x \leq 0.15$, a dome-shaped spin-singlet dimer phase below 130 K for $0.15 \leq x \leq 0.40$, and a magnetic short-range order for $x > 0.40$. A dimerization as similarly observed in α-RuCl$_3$ under high pressure occurs in the spin-singlet phase. It is suggested that Ru–Os pairs in the solid solutions tend to form dimers with short bonds and trigger the first-order transition in the presence of pseudo-threefold rotational symmetry for dimerization around a substituted Os atom only at low substitutions. This is a rare example of molecular orbital crystallization induced by elemental substitution in a highly disordered system. The short-range order at high substitutions may be related to a random-singlet state stabilized by bond disorder in the honeycomb net.

## 1. Introduction

After the proposal of a theoretical model by Kitaev in 2006[1] and the suggestion of material realization by Jackeli and Khaliullin in 2009,[2] the "Kitaev compound", which potentially gives an experimental platform for investigating the Kitaev spin liquid (KSL), has been explored and studied extensively.[3,4] Candidates are found in compounds with honeycomb nets made of transition metal ions in the $d^5$ electron configuration carrying the $j_{eff}$ = 1/2 state in the limit of strong spin–orbit interaction (SOI); iridates such as Na$_2$IrO$_3$ and α-Li$_2$IrO$_3$, and α-RuCl$_3$ are good examples.[5-8] Unfortunately, however, the KSL has remained elusive, because there are always cumbersome non-Kitaev interactions in actual compounds, which force a magnetic long-range order (LRO) instead of the KSL. Recent thermal conductivity measurements on α-RuCl$_3$ found fractional Majorana excitations when the LRO was suppressed under a magnetic field, which may be direct evidence of KSL, although the original KSL should appear only in zero magnetic field.[3,9]

α-RuCl$_3$ crystallizes in a layered structure with a honeycomb net made of Ru$^{3+}$ ions in the $4d^5$ electron configuration. The room-temperature structure is slightly distorted into a monoclinic $C2/m$ structure, while the monoclinic distortion seems to be removed upon cooling in a rhombohedral $R\bar{3}$ structure.[6,10,11] The compound exhibits a magnetic LRO with a zigzag spin arrangement at $T_N \sim 7$ K in high-quality crystals and a second transition at 14 K in crystals of lesser quality containing stacking faults.[7,10,12,13] Under high pressure, the LRO is replaced by a spin-singlet order above 0.2 or 1.7 GPa, which features a strong dimerization of Ru–Ru bonds.[14-16]

The Os$^{3+}$ ion ($5d^5$) is isoelectronic to the Ru$^{3+}$ ion and seems to have a larger SOI owing to heavy $5d$ electrons. Thus, it may be more advantageous to realize the KSL. However, previous trials to synthesize the stoichiometric "OsCl$_3$" ended in failure, resulting in an Os-deficient phase of Os$_x$Cl$_3$ ($x = 0.83$ or 0.81).[17,18] It crystallizes in a layered structure similar to that of α-RuCl$_3$, but consists of nanodomains of honeycomb nets with a highly disordered arrangement of Os ions.[18] Thus, it is unclear whether the compound harbors the Kitaev physics.

A dimerization that breaks the translational or rotational symmetry of a lattice is often observed in spin systems. Peierls instability is a well-known mechanism leading to a dimerization that typically occurs in one-dimensional systems such as CuGeO$_3$.[19] The bond modulations observed in CuGeO$_3$ are relatively small, ~1%.[20] Another type of dimerization is found in a valence bond crystal (VBC), mostly in which a two-dimensional lattice is covered with dimers in a specific pattern.[21,22] Bond modulations in the VBC are also expected to be small, because the underlying mechanism basically assumes a large repulsive Coulomb interaction that does not favor the shortening of bonds. In sharp contrast, there are many transition metal compounds that exhibit dimerizations with large bond modulations of 10–20%.[23] For example, rutile-related crystals exhibit large dimerizations in their metal chains: VO$_2$ (the bond alternation is 16%) and NbO$_2$ (20%) in their low-temperature phases below metal–insulator transitions, and metallic MoO$_2$ (21%) and WO$_2$ (22%) in the entire temperature range. These dimerizations have been understood in terms of the molecular orbital crystal (MOC):[23] a dimerization should occur when an energy gain by generating new chemical bonds between two metal ions with a pair of electrons in the bonding orbital exceeds an energy loss caused by the accompanied lattice deformation. Note that such a covalent chemical bond is always attractive so that a short bond is favored in this mechanism with minimal electron correlations.

Here, we report on the chemical substitution of Os for Ru in $Ru_{1-x}Os_xCl_3$; there is no such experiment thus far to the best of our knowledge. We observed an enhancement of antiferromagnetic interactions with increasing $x$ and a first-order transition to a spin-singlet dimer phase below 130 K at $0.15 \leq x \leq 0.40$ after a magnetic LRO was suppressed. We suggest that Ru–Os pairs generated in the random distribution of Ru and Os atoms in the honeycomb net prefer dimerization and stabilize the uniform spin-singlet state over the entire honeycomb lattice. This is a rare example of molecular orbital crystallization induced by elemental substitution. For $x > 0.40$, a magnetic short-range order (SRO) survives, which can be a candidate of the random-singlet spin liquid state.[24-27]

## 2. Experiments

Single crystals of $Ru_{1-x}Os_xCl_3$ of 25 different compositions were prepared from the mixtures of powder samples of α-$RuCl_3$ and $Os_xCl_3$ by the chemical transport method. $Os_xCl_3$ was prepared as described previously,[18] and α-$RuCl_3$ was prepared similarly by reacting Ru metal in a $Cl_2$ atmosphere: 200 mg of Ru powder and 0.4 ml of $CCl_4$ were put in an evacuated silica tube of 200 mm length and 10 mm diameter, and the tube was heated to 773 K for 100 h. An appropriate mixture of the two components of 500 mg weight in total was placed in an evacuated silica tube of 250 mm length and 10 mm diameter, and the tube was heated in a temperature gradient such as 923–1023 K or 873–923 K for 98 h; a lower average temperature was selected for a larger Os content. Hexagonal, thin platelike crystals of 0.1–0.5 mm size in the plane or aggregates of small crystals were obtained at the low temperature side of the tube. A film of Os metal was also produced near the middle of the tube, indicating that the chemical compositions of the obtained crystals were poor in terms of the Os content compared with the nominal composition. The actual $x$ values were determined to be 40–60% less than the nominal composition by energy-dispersive X-ray (EDX) analysis (JSM-IT100). The maximum $x$ value we could prepare was 0.67 starting from 90% Os composition. Unlike $Os_xCl_3$, no metal deficiency was detected for all the solid solutions.

The thus-obtained samples were characterized by powder X-ray diffraction (XRD) experiments. Two X-ray sources were used: a Cu–K$α_1$ radiation selected by a Johansson monochromator in a conventional diffractometer (Rigaku SmartLab) and a synchrotron radiation with a wavelength $λ$ of 68.4958 pm (18.1 keV) at BL08B in Photon Factory (PF). The former experiments were carried out at 4–300 K using powdered samples of $x$ = 0.23 and 0.33 in a reflection mode, and the latter was performed at 100–300 K using gently crushed crystals of $x$ = 0.23 and 0.58 sealed in a silica capillary of 0.3 mm diameter in a transmission mode. A complete powder XRD pattern was difficult to obtain, because the crystal was easily cleaved to show a strong tendency for the preferred orientation and also because stacking faults were inevitably introduced during the experimental setup. Electron diffraction (ED) experiments were carried out in a JEOL JEM–2010F with an acceleration voltage of 200 kV. A crushed powder sample of $x$ = 0.33 was used.

Magnetic susceptibility was measured in a Quantum Design MPMS-3, and heat capacity measurements were carried out in a Quantum Design PPMS. Raman scattering experiments were performed using a Nanobase XperRAM S series confocal Raman spectrometer equipped with a 532 nm Nd:YAG laser. The laser beam was focused to a spot diameter of about 3-4 μm on a single crystal through a microscope objective lens of 40× magnification, and the scattered light was detected via a holographic ultralow-frequency notch filter set. Temperature-dependent experiments between 5 and 300 K were carried out in a continuous He-flow cryostat.

## 3. Results
### 3.1 Sample characterization

All the XRD patterns of crushed crystals of $0 \leq x \leq 0.67$ were nearly identical at room temperature except for the effect of the preferred orientation, which clearly showed that solid solutions between Ru and Os were obtained in the wide composition range.

Figure 1 shows typical patterns for $x$ = 0.23 and 0.58, which are not similar to the simulated one for the rhombohedral $R$–3 structure, but for the monoclinic $C2/m$ structure. Note that, compared with the simulation, the series of peaks next to the sharp (0 0 1) peak on the high-angle side appear as a broad "continuum" in the experimental pattern, which is due to the high density of stacking faults already in the crystals or produced during sampling, as in the case of pure α-$RuCl_3$;[28] (0 0 $\ell$)-type peaks appear sharp as they originate from diffraction by the layers, while those including nonzero $h$ and $k$ indices are smeared by the random lateral sliding of the layers. Because of this, it was difficult to determine the lattice constants reliably. We carefully analyzed the $x$ = 0.58 data and obtained $a$ = 0.5991474(6) nm, $b$ = 1.040608(1) nm, $c$ = 0.6076631(6) nm, and $β$ = 109.182(1)° at 300 K, which are to be compared with those of α-$RuCl_3$, namely, $a$ = 0.59762(7) nm, $b$ = 1.0342(1) nm, $c$ = 0.6013(1), and $β$ = 108.87(2)°.[12] The unit cell volumes are 0.35783 and 0.35166 $nm^3$, respectively, larger by 1.8% for the solid solution, possibly reflecting the larger ionic radius of $Os^{3+}$ than of $Ru^{3+}$; the ionic radius of $Os^{3+}$ is not known because of the absence of Os(III) compounds, but should be larger owing to the expanded 5$d$ orbitals.

The electron diffraction pattern taken at room temperature for $x$ = 0.33 in Fig. 1(b) shows a pseudo-hexagonal symmetry with an inner set of six diffraction spots corresponding to the honeycomb lattice of metal atoms. Since these diffraction spots are sharper and more intense than those of $Os_xCl_3$,[18] there are no such nanodomains as in $Os_xCl_3$, which is consistent with the absence of metal deficiency. Thus, solid solutions preserving the crystal structure of the parent compound at room temperature have been successfully prepared.

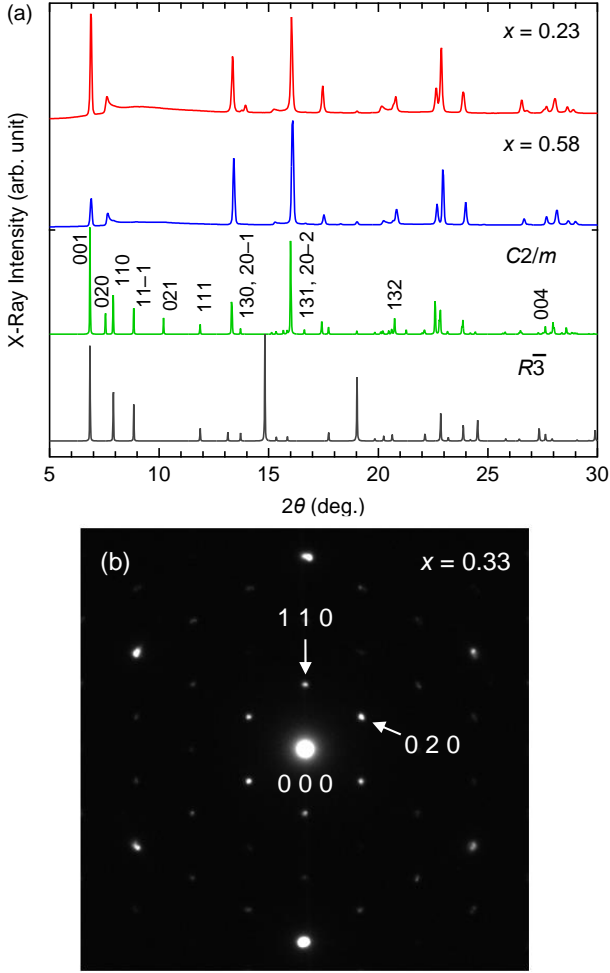

Fig. 1. (Color online) (a) Powder X-ray diffraction patterns of Ru$_{1-x}$Os$_x$Cl$_3$ samples with $x = 0.23$ (top) and 0.58 (second) recorded at BL08B in PF ($\lambda = 68.4958$ pm). Corresponding simulated patterns are shown for the $C2/m$ (third) and $R\bar{3}$ structure models (bottom). The indices of major diffraction peaks are given for the $C2/m$ structure. (b) Electron diffraction pattern from the $x = 0.33$ sample with the incident electron beam perpendicular to the metal layers. The indices of diffraction spots are based on the $C2/m$ structure.

### 3.2 Magnetic properties

Figure 2 shows the magnetic susceptibility $\chi$ of Ru$_{1-x}$Os$_x$Cl$_3$; an aggregate crystal was used for each measurement so that there was no specific direction for the applied field. The $\chi$ of the parent α-RuCl$_3$ shows a Curie–Weiss (CW) increase upon cooling, followed by a broad hump at 10–15 K (hardly visible in Fig. 2) and a sharp cusp at 8.5 K; it is a typical temperature dependence for a crystal including a small amount of stacking faults.[6] Thus, a magnetic LRO occurs at $T_N = 8.5$ K in a clean part of the crystal.

Similar cusps are observed at 12.5 and 12.0 K for 4% and 14% substitutions, respectively, and at 8.5 K for 15%. For 14% and 15%, in addition, there are gradual variations in $\chi$ with thermal hysteresis between the heating and cooling curves before the cusps. Then, for 19%, there is a sharp rise in $\chi$ at 127 K at the midpoint of the heating curve and a drop at 116 K upon cooling, indicative of a first-order phase transition. We define the transition temperature $T_d$ at the midpoint of the transition in the heating curve. Below the transition, the temperature dependence of $\chi$ is much more subtle than the CW behavior above $T_d$, and the $\chi$ value of $1.32 \times 10^{-4}$ cm$^3$ mol$^{-1}$ at 85 K is small [Fig. 2(b)]. Therefore, it is likely that a transition to a spin-singlet state occurs at $T_d$. The overall features of $\chi$ resemble those observed for α-RuCl$_3$ under high pressures above 0.2 GPa,[14] which has been ascribed to a transition to a spin-singlet phase with a strong dimerization of Ru–Ru bonds.[14,16]

As $x$ further increases, the $T_d$ transition in $\chi$ remains at similar temperatures and eventually disappears above $x = 0.45$. On the other hand, a gradual downturn is observed at a higher temperature of 190 K for $x = 0.26$ before the transition, which becomes more apparent at 200 K for $x = 0.33$ and 220 K for $x = 0.40$; the second downturn below 100 K for $x = 0.40$ must be a trace of the $T_d$ transition. The downturn is also discernible as a broad hump for $x = 0.58$ or 0.67, which is followed by a large Curie-like increase at low temperatures. We call this crossover temperature $T^*$.

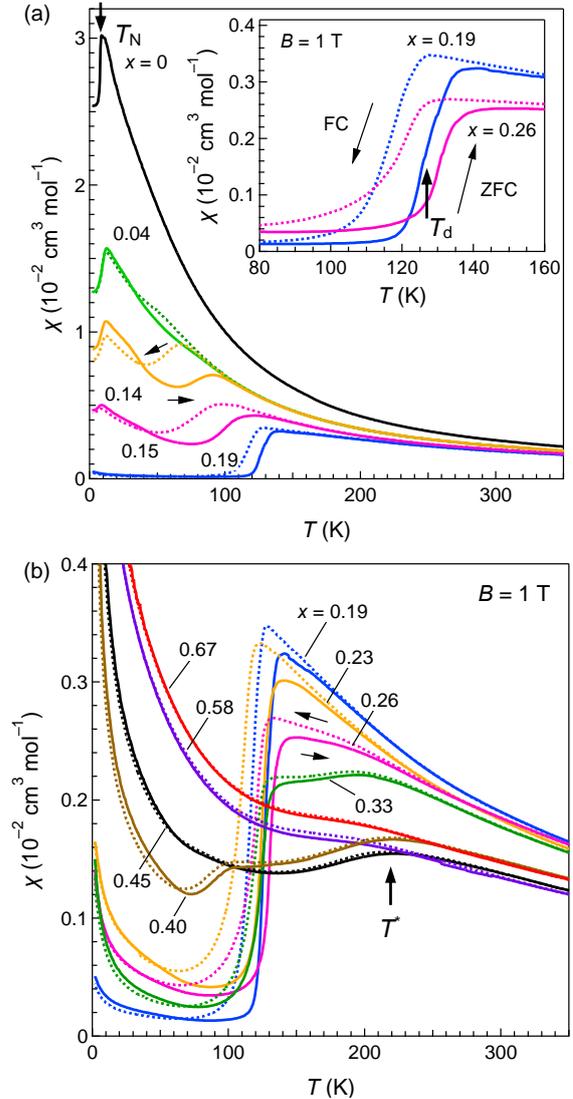

Fig. 2. (Color online) Magnetic susceptibilities of aggregate crystals of Ru$_{1-x}$Os$_x$Cl$_3$ with (a) $x \leq 0.19$ and (b) $0.19 \leq x \leq 0.67$;

data for selected compositions are shown for clarity. For each sample, a heating curve measured at a magnetic field of 1 T after zero-field cooling (ZFC) is shown by a solid line, and the following cooling curve in the same field (FC) is shown by a broken line. The inset of (a) expands the temperature range around the dimer transition for $x = 0.19$ and $0.26$, showing a large thermal hysteresis characteristic of a first-order transition. Three characteristic temperatures are marked: $T_N$, $T_d$, and $T^*$ related to the magnetic LRO, the spin-singlet dimer phase, and the magnetic SRO, respectively.

All the magnetic susceptibilities at high temperatures follow the CW law, $\chi = C/(T - \Theta_{CW})$, where $C$ is the Curie constant and $\Theta_{CW}$ is the Weiss temperature, as clearly evidenced by the inverse $\chi$ plot in Fig. 3(a). The inverse $\chi$ moves upward almost in parallel with increasing $x$, indicating that $\Theta_{CW}$ gradually decreases while keeping $C$ (thus the effective magnetic moment $\mu_{eff}$) almost unchanged. In fact, as plotted in Fig. 3(b), $\Theta_{CW}$ is positive at 40 K for $x = 0$, decreases to zero at $x \sim 0.15$, and becomes negative at $-130$ K at $x = 0.67$, whereas $\mu_{eff} \sim 2.3 \mu_B$ remains unaltered irrespective of substitutions ($\mu_B$ is the Bohr magneton). The CW parameters reported for single-crystal α-RuCl$_3$ are ($\Theta_{CW}$/K, $\mu_{eff}/\mu_B$) = (37, 2.14) for a magnetic field parallel to the honeycomb layer[13] and (40, 2.14) or (23, 2.25) for powder samples.[29-31] These $\mu_{eff}$ values are similar to those in various Ru(III) complexes.[32]

The fact that the $\mu_{eff}$ values per Ru$_{1-x}$Os$_x$ are nearly constant over the solid solutions may be reasonable, considering the same $d^5$ electron configuration of Ru$^{3+}$ and Os$^{3+}$ ions in the low spin state or in the $j_{eff} = 1/2$ state. However, it is rather surprising as it means that SOIs are also comparable for the two ions in contradiction to the naive expectation; the enhancement in $\mu_{eff}$ from $1.73 \mu_B$ for the ideal $j_{eff} = 1/2$ state may be attributed to additional effects such as the admixture of different electron configurations and the trigonal crystal field, as in α-RuCl$_3$.[33] On the other hand, the large variation in $\Theta_{CW}$ means that the ferromagnetic interaction (Kitaev interaction) present in the parent compound is reduced or cancelled by additional antiferromagnetic interactions that increase with $x$. The former originates from superexchange interactions via ~90° M–Cl–M bonds (M = Ru$_{1-x}$Os$_x$), and the latter from direct exchange interactions between metals. The observed dominance of the latter with increasing $x$ must be due to extended 5$d$ orbitals compared with 4$d$ orbitals: the Ru–Os and Os–Os pairs have larger antiferromagnetic interactions than the Ru–Ru pair in randomly substituted solid solutions. It is clear that this enhancement in antiferromagnetic interaction has stabilized the spin-singlet state for the solid solutions against KSL.

We estimate net magnetic interactions for Ru–Ru and Ru–Os pairs. In the mean-field theory, $\Theta_{CW}$ for a spin system with the nearest-neighbor magnetic interaction $J$ is given by $-zJS(S + 1)/(3k_B)$ (in the notation of the Hamiltonian given by $J \sum S_i \cdot S_j$), where $z$ is the number of nearest-neighbor sites and $k_B$ is the Boltzmann constant. Assuming spin 1/2, $\Theta_{CW}$ is equal to $-(3/4)J/k_B$ for a honeycomb lattice with $z = 3$. From $\Theta_{CW} = 40$ K for $x = 0$,

$J$(Ru–Ru)/$k_B$ is $-53$ K. Since one substituted Os atom loses three $J$(Ru–Ru) bonds and adds three $J$(Ru–Os) bonds in the dilution limit, the initial slope of $\Theta_{CW}$ as a function of $x$, which is approximately $-420$ K from Fig. 3(b), is expressed by $3[J(Ru–Os) - J(Ru–Ru)](-3/4)/k_B$. Thus, one obtains $J$(Ru–Os)/$k_B \sim 130$ K. The saturating tendency of $\Theta_{CW}$ for $x > 0.5$ in Fig. 3(b) suggests a similar antiferromagnetic value for $J$(Os–Os)/$k_B$.

Finally, note that the previous studies on Cr and Ir substitution showed different composition dependences of the CW parameters: for Ru$_{1-x}$Cr$_x$Cl$_3$, the average $\mu_{eff}$ gradually increases to the spin-only value of $3.87 \mu_B$ for spin-3/2 Cr$^{3+}$ while keeping the values of $\Theta_{CW}$ almost constant;[34,35] for Ru$_{1-x}$Ir$_x$Cl$_3$, the dilution by nonmagnetic Ir$^{3+}$ ions causes no change in $\mu_{eff}$ per Ru and a small decrease in the positive $\Theta_{CW}$.[36]

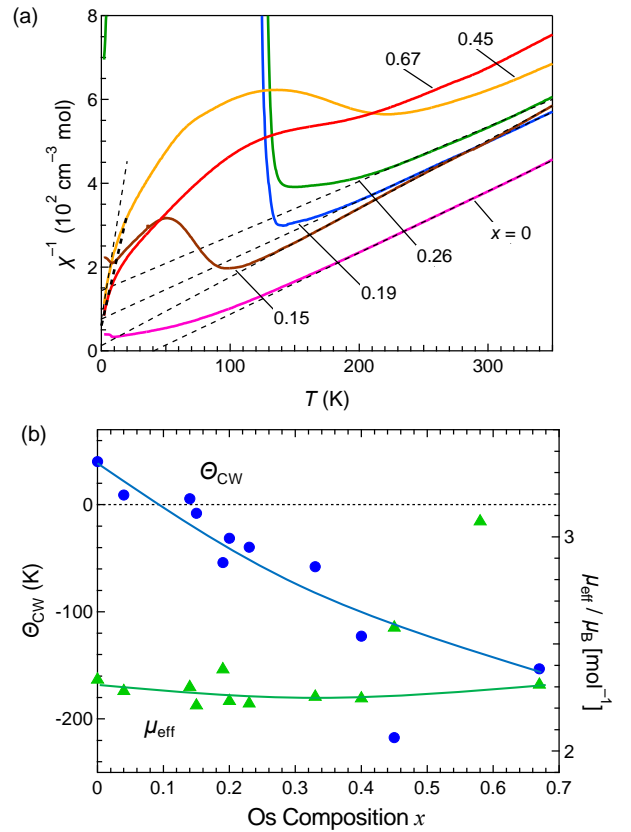

Fig. 3. (Color online) (a) Temperature dependences of inverse magnetic susceptibilities for selected compositions. The broken line on each data is a fit to the CW law without a temperature-independent term. (b) $x$ dependences of the CW temperature $\Theta_{CW}$ and effective magnetic moment $\mu_{eff}$ deduced from the CW fits at high temperatures such as that shown in (a). The curves on the data are guides to the eye. The data for $x = 0.45$ and $0.58$ considerably deviate from the curves, possibly because of the narrow temperature ranges for the fits that were limited by the presence of broad humps at $T^*$.

### 3.3 Phase diagram

The three characteristic temperatures determined by the magnetic susceptibility measurements are plotted in the $T$–$x$ phase diagram in Fig. 4. The antiferromagnetic LRO below $T_N$ survives up to $x \sim 0.15$ and is replaced by a spin-singlet phase (we call it the dimer phase) at $x = 0.15$–$0.40$ with the boundary of the dome shape below $T_d \sim 130$ K. The coexistence of the two phases at either end at $x \sim 0.15$ and $0.40$ is due to the first-order nature of the $T_d$ transition. Above $x \sim 0.40$, there is no LRO. A broad hump at $T^* \sim 200$ K appears above $x \sim 0.25$, suggesting that a magnetic SRO occurs at higher temperatures. For $x < 0.40$, the magnetic SRO is replaced by the dimer phase at low temperatures, while it remains as a ground state for $x > 0.40$.

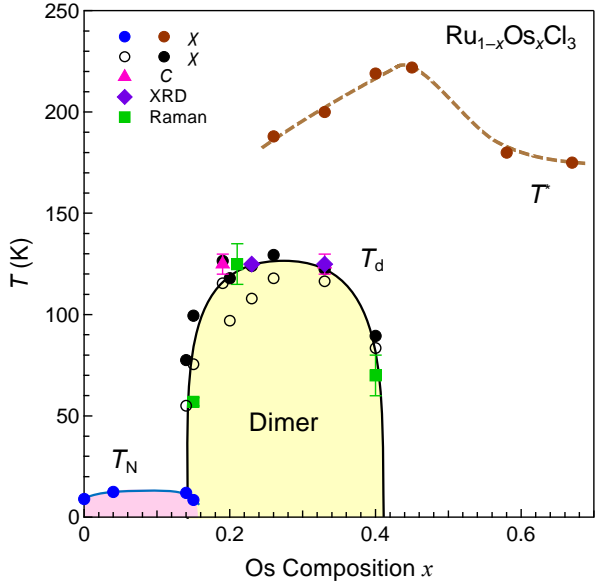

Fig. 4. (Color online) $T$–$x$ phase diagram of Ru$_{1-x}$Os$_x$Cl$_3$. The transition temperatures determined by magnetic susceptibility, heat capacity, XRD, and Raman scattering experiments are plotted as a function of the Os composition $x$: $T_N$, $T_d$, and $T^*$ related to the antiferromagnetic LRO at small $x$, the spin-singlet dimer order (dimer), and magnetic SRO, respectively. The solid and open circles in black represent $T_d$'s determined by magnetic susceptibility measurements upon heating and cooling, respectively. The lines on the data points are guides to the eye.

The phase diagram of Ru$_{1-x}$Os$_x$Cl$_3$ is unique and exceptional compared with those of other substitution systems studied thus far in the sense that a first-order transition emerges with substitution. In most cases, a certain LRO for a pure compound is suppressed with increasing doping and, near the quantum critical point, is replaced by a different LRO induced by a fluctuation associated with the parent LRO;[37] the transition to the second LRO should be of the second order. Otherwise, a glassy state appears or the parent LRO only fades away with increasing randomness induced by the substitution. It is plausible in Ru$_{1-x}$Os$_x$Cl$_3$ that the origin of the dimer phase is unrelated to the LRO of the parent compound.

We have measured the heat capacities of the $x = 0.19$, 0.33, and 0.45 samples. There is a peak at 120 K for the $x = 0.33$ sample, as shown in Fig. 5; a similar peak was observed at 120 K for $x = 0.19$ (not shown). These temperatures coincide with $T_d$, as plotted in the phase diagram of Fig. 4. Therefore, the $T_d$ transition is a thermodynamic phase transition in bulk. In contrast, the $x = 0.45$ sample showed no peak indicative of a phase transition at 2–300 K. Note that the $C/T$ curves of the two compounds coincide above the peak temperature of $x = 0.33$, whereas the $x = 0.33$ data are significantly reduced from the $x = 0.45$ data below the peak temperature. This indicates that a first-order structural transition occurs at $T_d$ for $x = 0.33$, which has reduced the basic lattice contribution. Moreover, note that there is a residual value in $C/T$ towards zero temperature only for $x = 0.45$, which will be discussed later.

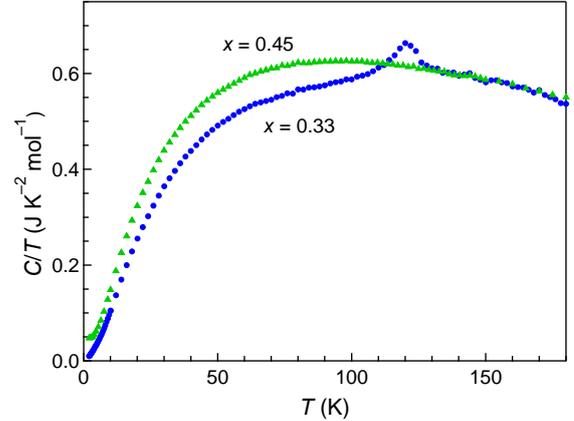

Fig. 5. (Color online) Heat capacity divided by temperature $C/T$ measured upon cooling for the $x = 0.33$ and 0.45 samples.

*3.4 Structural transition at $T_d$*

We have examined a possible structural change at $T_d$ in the temperature-dependent powder XRD experiments for $x = 0.23$. Figure 6 shows the temperature evolutions of XRD patterns in the ranges of diffraction angles including the (0 0 1), (0 0 3), and (0 0 4) diffraction peaks based on the $C2/m$ structure. Upon cooling from 150 K, a new peak grows on the low-angle side of each fundamental reflection, which is absent at 130 K and becomes discernible at and below 120 K. The new peaks are not indexed as forbidden reflections of the original unit cell but as superlattice reflections: the one near the (0 0 1) reflection is indexed as (1/2 3/2 0), and the others near the (0 0 3) and (0 0 4) reflections are as its third and fourth multiples, respectively. These superlattice reflections correspond to a $2a \times 2a$ superstructure of the honeycomb lattice of metal atoms based on the $R\bar{3}$ structure. Similar superlattice reflections were observed below $T_d$ for $x = 0.33$, but not for $x = 0.58$. Therefore, the structural transition at $T_d$ concurs with the magnetic transition.

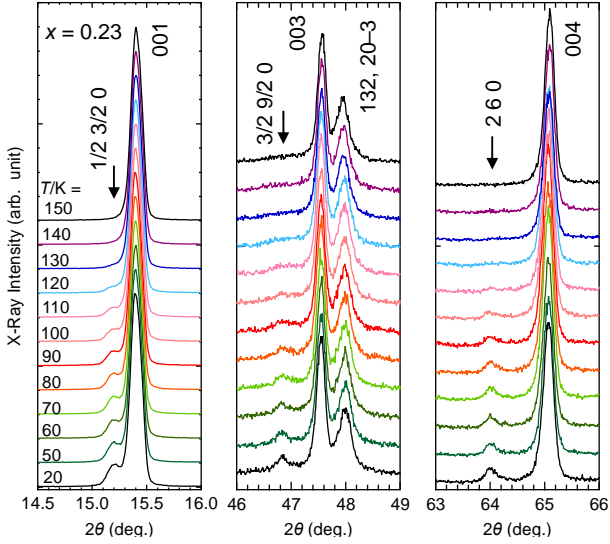

Fig. 6. (Color online) Temperature evolution of the powder XRD patterns measured upon heating using Cu–Kα1 radiation (λ = 154.05 pm) for the $x = 0.23$ sample. Three angle ranges including the (0 0 1) (left), (0 0 3) (middle), and (0 0 4) fundamental diffraction peaks (right) are shown. In each figure, a new peak marked by the arrow on the low-angle side grows upon cooling below 120 K. All the indices assume the monoclinic $C2/m$ structure.

Clear evidence of a structural transition at $T_d$ has been obtained by Raman scattering experiments using small single crystals with five compositions of $x = 0.04$, 0.15, 0.21, 0.40, and 0.58 (Fig. 7). First, we note that all the Raman spectra at 300 K resemble that of α-RuCl$_3$,[16,38] indicating that the $C2/m$ structure is preserved for all the solid solutions. According to the previous Raman experiments on α-RuCl$_3$,[38] there are two Raman active modes, A$_g$ and B$_g$, based on the $C2/m$ structure, which are almost degenerate in energy and are distinguished depending on polarization conditions. In our Raman spectra obtained using unpolarized light, both A$_g$ and B$_g$ modes were observed. In pure α-RuCl$_3$, there are six A$_g$/B$_g$ modes at room temperature with energies of 14.5, 20.2, 27.7, 33.6, 36.9, and 42.3 meV;[38] for clarity, we call them A$_g$(i) with $i = 1$–6 in this order. Among them, four intense modes are apparently observed at 300 K for the $x = 0.04$ sample in Fig. 7(a): the corresponding A$_g$(1), A$_g$(2), A$_g$(4), and A$_g$(5) modes shown by the arrows are located at 14.1, 19.8, 36.2, and 38.2 meV, respectively; slight reductions in energy are due to the heavy mass of the Os atom. Upon cooling, they show a small hardening and remain present down to 5 K; a structural transition observed in the previous experiments for α-RuCl$_3$ is missing for $x = 0.04$.[11,38]

For $x = 0.15$, in sharp contrast, all those peaks suddenly disappear at ∼60 K, replaced by a completely different Raman spectrum below the temperature. A similar behavior is observed for $x = 0.21$ at 120 K and for $x = 0.40$ at ∼70 K. On the other hand, such a change is absent for $x = 0.58$, and the 40 meV mode smoothly gains its intensity down to 5 K. The transition temperatures thus determined are plotted in the phase diagram of Fig. 4, which exactly agree with the phase boundary of the dimer phase determined by other measurements.

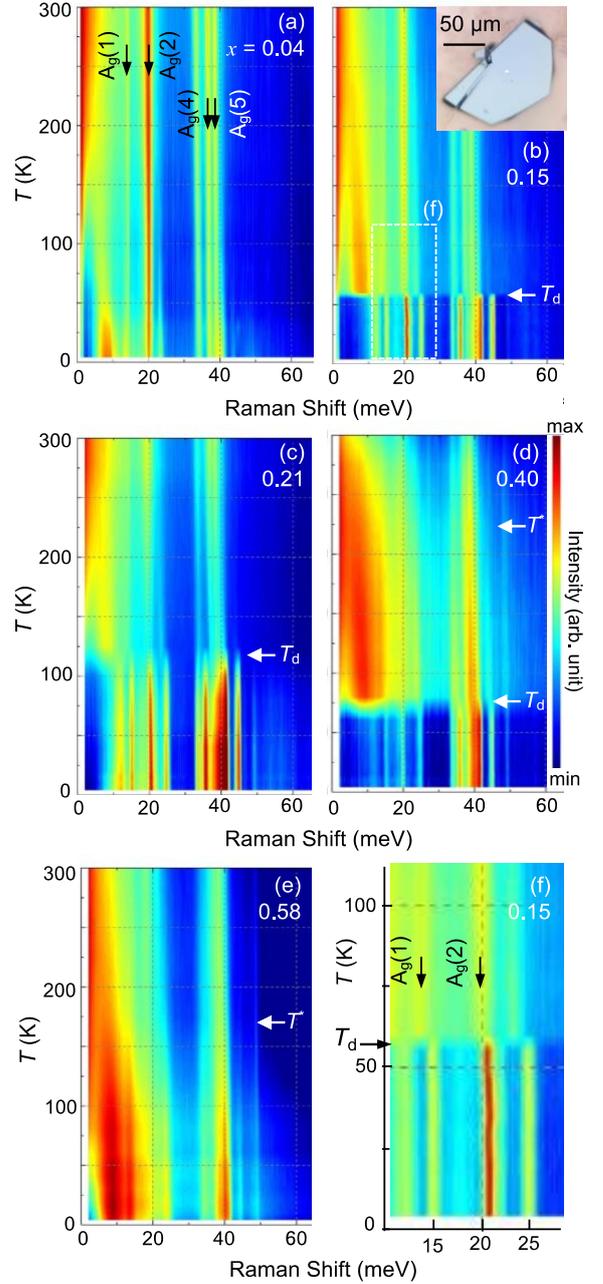

Fig. 7. (Color online) Temperature evolution of the Raman scattering data measured upon cooling for $x = 0.04$ (a), 0.15 (b), 0.21 (c), 0.40 (d), and 0.58 (e). The inset in (b) at the top right shows a photograph of the $x = 0.15$ crystal used for measurements. The area surrounded by the broken rectangle in (b) is expanded in (f). Four A$_g$ modes marked by the arrows in (a) are analogies based on the previous Raman scattering experiments on α-RuCl$_3$ at room temperature.[16,38] $T_d$ is defined as a temperature with a discontinuous change in the Raman spectrum, and $T^*$ determined by magnetic susceptibility is shown by the arrow.

Let us look at the changes in the spectrum at $T_d$ in more detail for $x = 0.15$ near the phase boundary. As shown in the expansion of Fig. 7(f), the A$_g$(2) mode located at 20.0 meV jumps to 22.5 meV below $T_d$. Moreover, the A$_g$(1) mode at 13.6 meV splits into two modes at 12.0 and 14.8 meV below $T_d$. Similar changes were observed in the high-pressure Raman scattering experiments throughout the $C2/m$ to $C2$ transition with dimerized Ru–Ru bonds above 1.7 GPa.[16] In particular, the energy shift of the $A_g(2)$ mode, which is associated with the in-plane relative movement of Ru atoms, was considered to be direct evidence of dimerization. Moreover, at the transition to the dimer phase under high pressure, the $A_g(4)$ and $A_g(5)$ modes at ~40 meV split into five modes with a large enhancement in the intensity. Corresponding changes are clearly observed for $x = 0.15$ in our case and may also occur for $x = 0.21$ and 0.40. However, for the latter samples, the phonons are significantly broadened, which hinders a clear thermal tracing.

Figure 8 shows the lowest temperature Raman spectra with those of α-RuCl₃ at ambient pressure[38] and a high pressure of 5 GPa.[16] Note that the spectra for $x = 0.15$, 0.21, and 0.40 in the dimer phase resemble that of α-RuCl₃ at 5 GPa, taking into account certain energy shifts caused by the Os substitution and the lattice contraction under pressure. Therefore, it is plausible that the spin-singlet phase of $Ru_{1-x}Os_xCl_3$ has a similar dimerized structure with short and long M–M bonds as realized under high pressure. On the other hand, the $x = 0.58$ spectrum is different, particularly regarding the lack of the intense 41-meV peak, and is more similar to the $x = 0.04$ and 0 spectra, ignoring the peak broadening by enhanced disorder. Although it is difficult to argue about the exact structure of the $x = 0.58$ sample based only on the Raman data, this change in trend may signal a gradual reversal towards the initial α-RuCl₃ structure.

Another interesting observation in Fig. 8 is the presence of broad peaks centered at 8–9 meV only for $x = 0.04$ and 0.58, which seem to appear upon cooling below ~50 and ~200 K, respectively (Fig. 7). Note that a similar peak is present for $x = 0.40$ below ~300 K but disappears at $T_d$. These low-energy excitations for $x = 0.40$ and 0.58 must correspond to a magnetic SRO, because they grow approximately below $T^*$ and their lineshapes are distinct from those of other phonons. On the other hand, the origin of the broad peak for $x = 0.04$ is unclear, although it may also be magnetic. Note that its intensity is much lower than that for $x = 0.58$, as compared in the original intensity profiles shown in Fig. 7.

### 3.5 Disordered phase at high substitutions

There is no LRO above $x = 0.40$ beyond the dimer phase, and only a magnetic SRO seems to develop below $T^*$. The SRO is already present for $0.23 \leq x \leq 0.40$ but replaced by the dimer transition upon cooling. On the other hand, the SRO persists down to $T = 0$ for a larger $x$. Although there is no structural transition at $T^*$, we have detected a characteristic lattice deformation at around $T^*$. Figure 9 shows a temperature-dependent XRD pattern for $x = 0.33$, in which the (0 0 3) peak moves to the high-angle side down to 240 K and to the low-angle side below 220 K, while the (1 3 2) peak moves to the high-angle side continuously; we observed a similar variation at around $T^*$ for $x = 0.58$. This means that the interlayer distance starts to increase upon cooling at around $T^*$ with the intralayer distance smoothly decreasing. Thus, the magnetic SRO may cause an anisotropic lattice deformation and a negative thermal expansion only for the interlayer spacing.

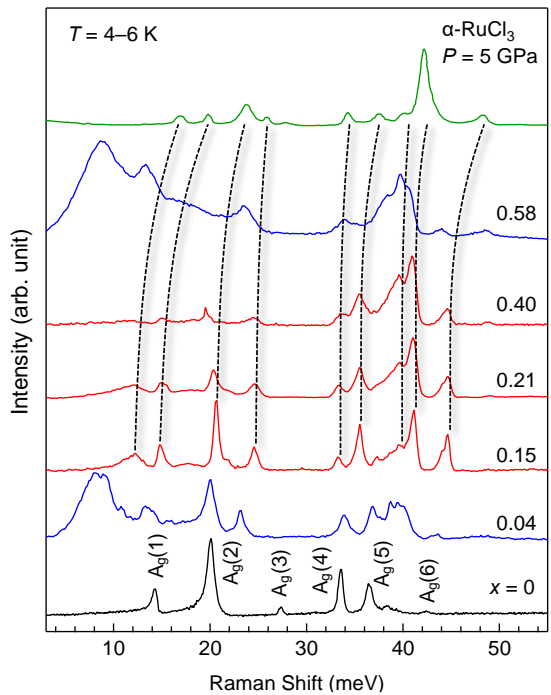

Fig. 8. (Color online) Raman spectra of the solid solutions at low temperatures of 4–6 K. Those of α-RuCl₃ at ambient pressure (bottom)[38] and 5 GPa (top)[16] are also shown for comparison. The vertical broken lines indicate approximate correspondence between the peaks in the dimer phase and the high-pressure data.

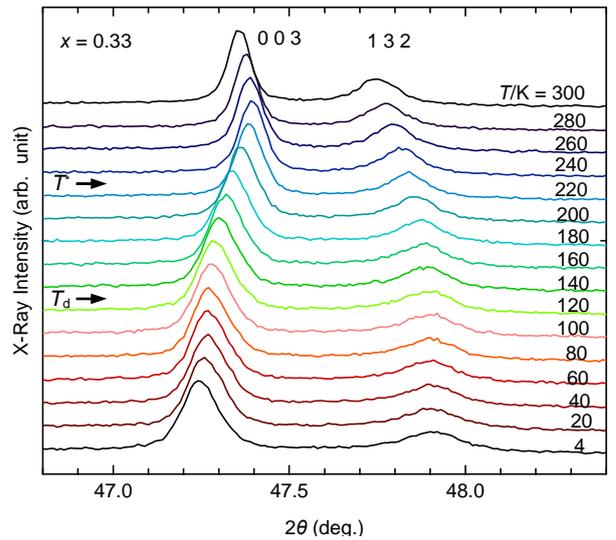

Fig. 9. (Color online) Temperature evolution of the powder XRD patterns obtained using Cu–Kα1 radiation for $x = 0.33$, which includes the (0 0 3) and (1 3 2) diffraction peaks.

The magnetic susceptibilities of the large-$x$ samples exhibit divergent increases at low temperatures, as shown in Fig. 2(b). The divergent contribution tends to mask the broad humps at $T^*$. Such a low-temperature increase is often observed in low-dimensional and/or frustrated quantum spin systems and is called the "Curie tail".[39-41] It originates from weakly interacting unpaired spins that lost their partners in singlet-based ground states or simply become nearly free from the surrounding spins at crystalline defects. In particular, in a dimer crystal with a spin-singlet ground state, an unpaired spin is generated as an orphan spin when the counter ion is replaced by a nonmagnetic impurity ion. A typical example is found in the dimer phase of $VO_2$, in which the number of orphan spins scales with the number of nonmagnetic substituents such as $Ti^{4+}$ ions.[23,42] Since the present compound is also related to spin-singlet dimers and is a chemically substituted system, the Curie tail must have a similar origin.

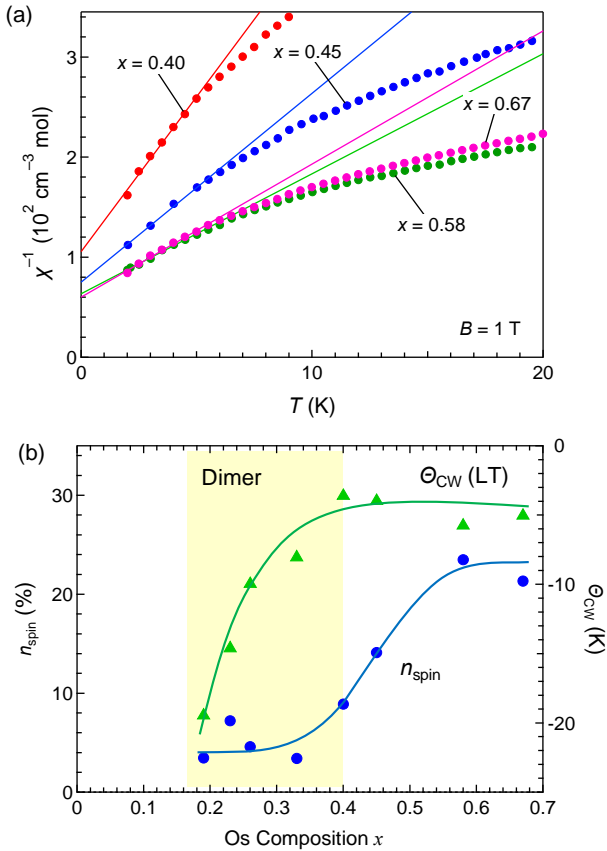

Fig. 10. (Color online) (a) Inverse magnetic susceptibilities of the $x$ = 0.40, 45, 0.58, and 0.67 samples. The solid line on each dataset is a fit to the CW form in a $T$ range of 2–5 K, from which deduced are the number of orphan spins $n_{spin}$ and the Weiss temperature $\Theta_{CW}$(LT) plotted in (b) as a function of $x$.

We fit the low-temperature inverse $\chi$ data at 2–5 K to the linear form, as shown in Fig. 10(a). The fits should approximately estimate the number of orphan spins and interactions felt by them, because the Curie component becomes dominant at such low temperatures compared with other ignorable less-$T$-dependent components. The slope decreases with increasing $x$, indicative of more orphan spins generated. From the thus-obtained Curie constants, the number of free spins $n_{spin}$ is calculated assuming the entity of spin 1/2 with the Landé $g$ factor of 2. As shown in Fig. 10(b), $n_{spin}$ is relatively small ~5% in the dimer phase and significantly increases for $x > 0.40$ to saturate at ~23%. The former small contribution must originate from structural defects in the dimer crystal, and the latter may be from the SRO, in which nearly a quarter of spins behave as orphan spins. On the other hand, the Weiss temperature changes from –20 K for $x$ = 0.20 to –5 K for $x$ = 0.67, much smaller than those from the high-temperature CW fits. Therefore, the ground state at high substitutions is significantly disordered after the suppression of the dimer structure. We will return to the origin of the SRO in the discussion.

The "intrinsic" magnetic susceptibility is estimated by subtracting the low-temperature CW component, as shown for $x$ = 0.45 and 0.58 in Fig. 11. The broad peak at $T^*$ is now apparent for each sample, and $\chi$ seems to approach a large residual value toward $T$ = 0 and finally drops to zero; the downturn below ~50 K is an artifact caused by CW fitting ignoring a temperature-independent term. Note that the absolute value is not reliable, because a large contribution from orphan spins has been simply subtracted.

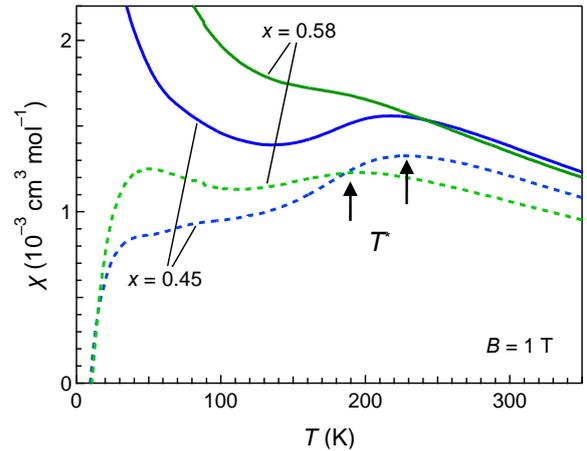

Fig. 11. (Color online) Magnetic susceptibilities of the $x$ = 0.45 and 0.58 samples and their corrections (broken lines) after the subtraction of the corresponding low-temperature CW components.

Figure 12 shows the low-temperature heat capacity measured under various magnetic fields for $x$ = 0.45. The zero-field data show a linear dependence in the $C/T$ versus $T^2$ plot of the inset below ~10 K, followed by a weak upturn below ~5 K. From the linear fit, we estimate the $T$-linear component in $C$ as 33.9(5) mJ K$^{-2}$ mol$^{-1}$ and a Debye temperature of 219 K. In contrast, the $x$ = 0.33 sample has a much smaller $T$-linear component than that for $x$ = 0.45: in fact, a similar fit yielded a much smaller value of 3.9(2) mJ K$^{-2}$ mol$^{-1}$ for $x$ = 0.33. On the other hand, the application of magnetic fields markedly changes the $T$-dependence of $C/T$: the upturn toward 2 K is enhanced with increasing field first and then a broad maximum appears with the peak top moving to high temperatures with further increasing field. Such Schottky-like behavior usually originates from disordered spins. Thus, the observed $T$-linear component in $C$ is due to the residual spin entropy associated with the SRO. On the origin of the $T$-linear heat capacity, one may consider a spin glass state. However, this is unlikely because the magnetic susceptibilities for $x \geq 0.45$ in Fig. 2(b) are different from those of a typical spin glass with a broad peak followed by an opening of thermal hysteresis upon cooling.[43]

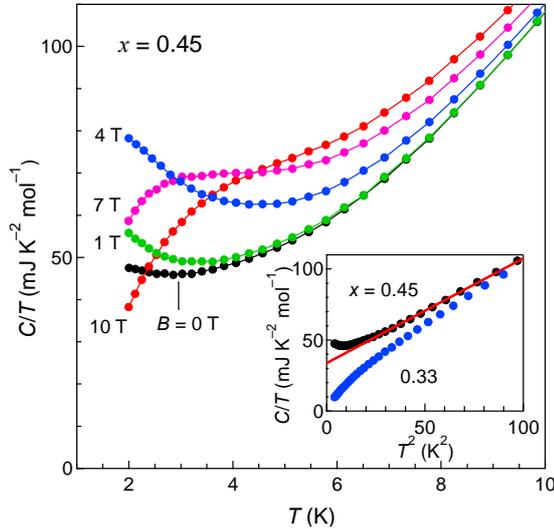

Fig. 12. (Color online) $C/T$ at low temperatures under various magnetic fields for $x = 0.45$. The inset plots $C/T$ as a function of $T^2$ with a linear fit shown by the red line.

## 4. Discussion

### 4.1 Dimer crystallization in ruthenates and other transition metal compounds

It is known that phase transitions to dimerized structures occur in some ruthenates. $Li_2RuO_3$ crystallizes in a layered structure containing a honeycomb net of $Ru^{4+}$ ($4d^4$) ions and exhibits a phase transition at 540 K to such a dimerized structure as depicted in Fig. 13, in which short and long Ru–Ru bonds alternate with a difference of ~8.5%.[44] On the other hand, $\alpha$-$RuCl_3$ with $Ru^{3+}$ ($4d^5$) shows a dimerization under high pressure to a different pattern covering the honeycomb lattice as also depicted in Fig. 13, in which the alternation of Ru–Ru bonds reaches 10%.[14] Note that the dimers are arranged in a staggered way in the former, whereas they are aligned in the same direction in the latter. The staggered pattern in $Li_2RuO_3$ was explained by taking into account a magnetoelastic coupling.[45] Another important difference between the two compounds is on the Ru–Ru bond lengths in the parent structures: the average bond lengths are much larger for $\alpha$-$RuCl_3$ (0.354 nm at ambient pressure)[10] than for $Li_2RuO_3$ (0.293 nm at 600 K),[44] reflecting the difference in ionic radius between the $Cl^-$ ion (0.181 nm) and the $O^{2-}$ ion (0.140 nm).

Let alone ruthenates, dimerizations with large bond alternations are ubiquitously observed in many transition metal compounds.[23] One related example is another Kitaev candidate, $\alpha$-$Li_2IrO_3$, in which a similar dimer arrangement as in $Li_2RuO_3$ occurs with 11% bond alternation in the honeycomb net of $Ir^{4+}$ ($5d^5$) under high pressure.[46] More examples are found in wide classes of crystal structures other than the honeycomb net. In addition to the rutile-related crystals mentioned in the introduction, $NbCl_4$ and $MoBr_3$ have quasi-one-dimensional structures with strong dimerizations of 22% and 11% in their metal chains, respectively. It is also the case for three-dimensional spinel compounds such as $MgTi_2O_4$ and $LiRh_2O_4$ with weaker dimerizations. All these dimerizations have been understood in terms of MOC.[23] Depending on the electron filling in the $d$ orbitals and parent crystal structure, dimers or even larger "molecules" such as trimers are produced in the rigid framework made of other chemical bonds, as in the case of $\pi$ electrons in organic compounds. When all itinerant $d$ electrons are trapped by the molecular orbital state, the transition is accompanied by a metal–insulator transition with marked changes in electrical conductivity and magnetic properties.

We think that the dimerizations in the ruthenates and $Ru_{1-x}Os_xCl_3$ are classified as MOCs rather than VBCs because of the large bond alternations and the relatively weak electron correlations. An interesting question is why the transition in $Ru_{1-x}Os_xCl_3$ occurs only in the intermediate range of compositions in a well-defined form despite the large substitutional disorder; there is no example of the MOC for such a solid solution system to the best of our knowledge.

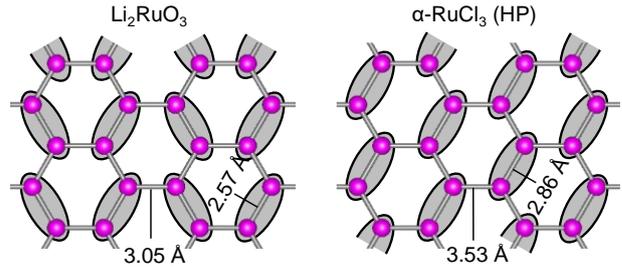

Fig. 13. (Color online) Schematic representations of dimer arrangements in $Li_2RuO_3$ (left)[44] and $\alpha$-$RuCl_3$ under high pressure (right).[14] The oval on a pair of atoms represents a dimer with a shorter bond.

### 4.2 Dimer crystallization in $Ru_{1-x}Os_xCl_3$

Dimer crystallization may be allowed in $Li_2RuO_3$ because of the short Ru–Ru bonds, whereas it is difficult for $\alpha$-$RuCl_3$ with long bonds at ambient pressure, because the loss in lattice energy necessary to make shorter bonds with direct chemical bonding is very large. Thus, it becomes possible only when the compound is squeezed under pressure. The transition occurs above 0.2 (or 1.9) GPa, and the transition temperature gradually increases from 120 to above 300 K at 1.5 GPa.[14] Hence, one may suspect that the Os substitution induces chemical pressure. However, the lattice of $\alpha$-$RuCl_3$ does not shrink but expands with the substitution of larger Os ions, pointing to a rather negative chemical pressure. Moreover, $T_d$ remains at ~130 K irrespective of the Os content and then disappears in $Ru_{1-x}Os_xCl_3$, in contrast to the continuous increase in transition temperature with increasing pressure in $\alpha$-$RuCl_3$. Thus, the chemical pressure effect must be irrelevant.

We focus on the fact that the dimer phase exists as a dome only at intermediate compositions in the phase diagram of Fig. 4. Neither local Ru–Ru nor Os–Os pairs must be helpful in stabilizing the dimer phase. This fact strongly suggests that the driving force is ascribed to the formation of local Ru–Os dimers; interestingly, the maximum $T_d$ is close to $J$(Ru–Os)/$k_B$ ~ 130 K estimated in Sect. 3.2. As evidenced by the results of the CW analyses of the high-temperature magnetic susceptibility, the $Ru^{3+}$ and $Os^{3+}$ ions possess similar electronic states with nearly equal magnetic moments of $2.3\mu_B$. The critical difference between them should be their effective size in a crystal: the 5$d$ orbitals of $Os^{3+}$ are more expanded than the 4$d$ orbitals of $Ru^{3+}$. Therefore, a Ru–Os pair can easily form a dimer by generating a direct metal–metal bond with less lattice deformation than a Ru–Ru pair (here, 'pair' and 'dimer' refer to M–M with the normal and shorter bonds, respectively). On the other hand, such a dimer may not be preferred for an Os–Os pair having already a considerable overlap between their extended 5$d$ orbitals, in which the energy gained by generating additional chemical bonds is relatively small. A question then is why the dome is not located around the equimolar solid solution, but shifted to the Ru side in the phase diagram; if Ru–Os dimers actually stabilize the dimer phase, the dome should be located around $x = 0.5$.

To understand what happens actually in this heavily substituted system with large configurational disorder, it is intuitive to illustrate possible arrangements of Ru and Os atoms in the honeycomb net, which are supposed to be "random" as determined during chemical reactions at high temperatures and then quenched. Figures 14(a) and 14(b) show them for $x = 0.20$ and 0.40, respectively. Since we observed a $2a \times 2a$ superstructure below $T_d$ and also from the resemblance in Raman spectra, the dimer arrangement realized in $Ru_{1-x}Os_xCl_3$ must be similar to that of $\alpha$-RuCl$_3$ under high pressure with all the dimers aligned in the same direction.[14]

First, let us consider a local arrangement with one substituted Os atom surrounded by three Ru atoms, as depicted in the top left of Fig. 14(a). There are three directions for the formation of Ru–Os dimers named 'H', 'R', and 'L'. All the three dimers are equally allowed unless another Os atom replaces one of the three Ru atoms. In this sense, the central Os atom possesses threefold rotational symmetry for dimerization. Assuming that the 'R' dimer is selected, all the nearby Ru–Os dimers can occur in the same direction to have a uniform arrangement of aligned dimers, as depicted in Fig. 14(a). We emphasize that this is possible only for a diluted case because of the presence of a threefold rotation axis at most Os sites. However, even when an Os–Os pair is created with increasing $x$, the formation of an R dimer is not disturbed unless the Os–Os pair lies in the same direction; Os–Os pairs in the other directions allow the formation of the uniform arrangement of R dimers. As a result, a well-defined structural transition with a large spatial coherence can occur despite the enormous configurational disorder. It becomes a first-order transition, probably because the energetic stabilization by dimerization is large, as commonly observed for most phase transitions associated with molecular orbital crystallization.[23] Hence, the present system gives us a unique dimer crystallization induced by elemental substitution.

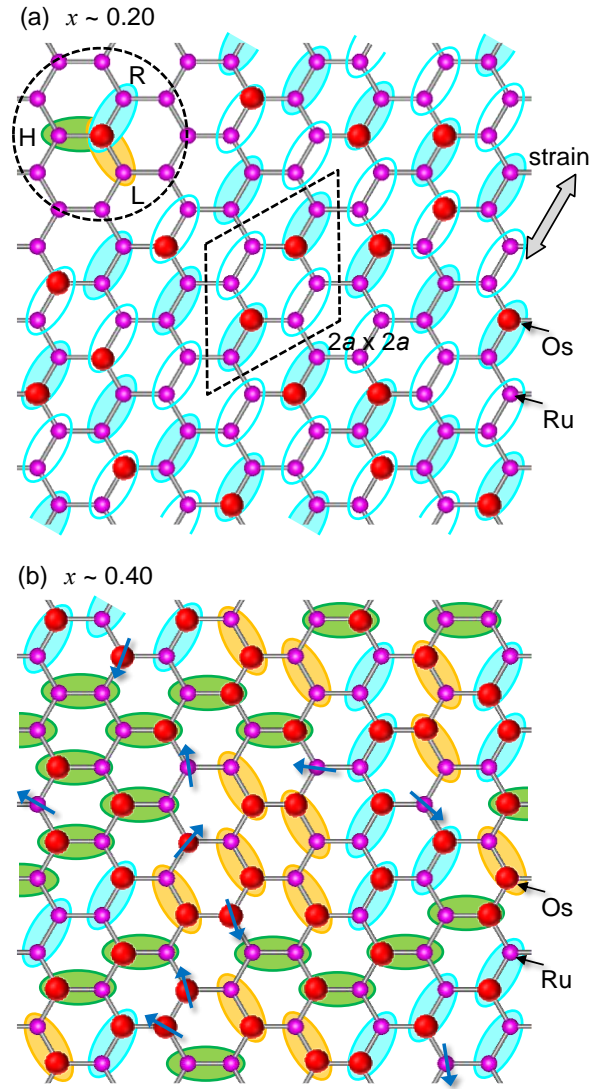

Fig. 14. (Color online) Schematic representations of dimer arrangements in the honeycomb net made of Ru and Os atoms in $Ru_{1-x}Os_xCl_3$. (a) Possible metal arrangement for $x = 0.20$, in which approximately 20% of Ru atoms (small balls) are randomly replaced by Os atoms (large balls). Shown inside the broken circle at the top-left corner is a local arrangement around one substituted Os atom in the dilution limit: one of the three Os–Ru dimers named 'H', 'R', and 'L' is evenly selected; there is threefold rotational symmetry for dimerization. The rhomboid represents a $2a \times 2a$ superlattice in the honeycomb net that assumes average atomic occupations and a possible dimer-bond modulation: the shaded and open ovals represent dimers with short and long bonds, respectively. (b) Possible metal arrangement for $x \sim 0.40$. The arrows represent orphan spins of about 10% appearing at the boundary of microscopic domains with a dimer order at low temperatures. Note that the actual states of $x > 0.40$ at $T = 0$ may not be described by this cartoon but may be dynamical states with spin singlet correlations related to the random-singlet state.[24-27]

The transition is driven by the Ru–Os dimerization and is allowed only for diluted compositions by the pseudo-threefold rotational symmetry on the honeycomb lattice. Apparently, the more the Ru–Os dimers, the more stable the dimer phase. An actual phase transition should occur when the density of Ru–Os dimers and thus the total chemical bonding energy gained by the dimerization increases with $x$ to exceed the critical value required to overcome the energy loss caused by the lattice deformation to the $2a \times 2a$ superstructure, which happens to be $x \sim 0.15$.

The observed $2a \times 2a$ superstructure, not the $a \times a$ structure as in α-RuCl$_3$ under high pressure, seems to indicate that the dimerization occurs at every other row perpendicular to the dimer direction. Then, one would expect that only half of the spins are involved in singlet states. However, since the observed magnetic susceptibility is completely nonmagnetic, all the spins should participate in singlet states in the dimer phase. One possible model assumes that two spin-singlet dimers with short (shaded ovals) and not-so-short bonds (open ovals) alternate in the dimer direction, as depicted in Fig. 14(a). It is considered that the $2a \times 2a$ superstructure is realized because of the low density of active Ru–Os dimers; the $a \times a$ superstructure may be preferred in α-RuCl$_3$ under high pressure with all active Ru–Ru dimers; the $2a \times 2a$ superstructure may be a compromise. Note that, in the uniform $2a \times 2a$ superstructure, not only Ru–Os but also Ru–Ru and Os–Os dimers occur in singlet states, although the chemical bonding energies of Ru–Ru and Os–Os dimers themselves are less than that of Ru–Os dimers; they are just forced to become dimers so as to maintain the uniform structure without an additional increase in lattice energy.

Next, we consider a possible situation at high substitutions such as $x = 0.4$, as illustrated in Fig. 14(b). Now, at such high substitutions, there are many Os–Os pairs or even trimers in the honeycomb net, so that the above-mentioned threefold rotational symmetry for dimer formation becomes incomplete. When an Os–Os pair occurs in the R direction, a Ru–Os dimerization is only possible in the L or H direction. Since the directions of Os–Os pairs are randomly quenched, one expects everywhere embryos or small domains with dimers aligned in either of the three directions, and none of them can grow large at low temperatures, in contrast to the case of low substitutions. As a result, such a microscopic domain mixture as depicted in Fig. 14(b) may be eventually attained. Thus, the dimer phase transition should be suppressed at high substitutions, which happens to occur for $x > 0.40$.

It is likely that at the boundary of domains, many orphan spins are generated on Ru/Os atoms, which cannot find partners to form singlets because of the structural incoherence. The observed high density of orphan spins of 23% for $x > 0.50$ suggests that "domain size" becomes very small with further increasing $x$. However, note that a more complex dynamical state with a singlet correlation may be a better description for the SRO, as will be discussed in the next section.

*4.3 Disordered state beyond the dimer phase*

After the dimer phase is suppressed at high Os substitutions above $x \sim 0.40$, only a magnetic SRO is observed below $T^*$. This must be related to antiferromagnetic or singlet correlations in the uniform honeycomb net at high temperatures in the absence of structural dimerization.

Upon cooling, such microscopic domains as illustrated in Fig. 14(b) tend to be generated with many orphan spins left at the boundary. However, the cartoon in Fig. 14(b) may not be appropriate to represent the actual ground states for $x > 0.40$, because there is a large magnetic susceptibility of $\chi_0 \sim 1 \times 10^{-3}$ cm$^3$ mol$^{-1}$ left after the subtraction of the orphan spin contribution in Fig. 11: most spins are not in singlets but still exhibit local moments. Moreover, the heat capacity has a large $T$-linear component, suggesting a gapless spin excitation (Fig. 12). Provided an analogy to the Fermi liquid, the corresponding magnetic susceptibility is calculated to be $4.7 \times 10^{-4}$ cm$^3$ mol$^{-1}$ for the Wilson ratio of 1, which is nearly half of $\chi_0$ (the Wilson ratio is about 2). In the Raman spectra shown in Figs. 7 and 8, on the other hand, the low-energy excitations disappear at $T_d$ for $x = 0.40$, while surviving down to 5 K with a (pseudo) excitation gap of a few meV. This suggests the formation of a certain dynamical singlet state. Therefore, it is considered that the ground state at high substitutions is a dynamically disordered spin state probably with singlet correlations in a structurally uniform honeycomb lattice with a large substitutional disorder. The singlet correlations must involve not only nearby pairs but also distant pairs, resulting in a nearly gapless state, as in the case of the long-range resonating valence bond state.[47]

A randomness-induced quantum spin liquid state has been proposed for frustrated triangular and kagome antiferromagnets with quenched disorder.[25] In addition, for a honeycomb spin system with frustrated nearest- and next-nearest-neighbor antiferromagnetic interactions ($J_1$ and $J_2$), a similar random-singlet state with gapless excitations is predicted to neighbor antiferromagnetic LRO and VBC phases when a certain randomness is introduced to the magnetic couplings.[26] Interestingly, this random-singlet state is expected to show a $T$-linear heat capacity at a low temperature and a broad hump followed by a Curie tail in magnetic susceptibility,[27] as observed for our system. Moreover, the Wilson ratio calculated is about 2,[48] in good agreement with our observation. In Ru$_{1-x}$Os$_x$Cl$_3$, frustration is neither induced by geometry nor by a competition between $J_1$ and $J_2$ but must be caused by a competition between ferromagnetic Ru–Ru bonds and antiferromagnetic Ru–Os and Os–Os bonds. In addition, randomness should arise from bond disorder caused by the random Ru/Os occupation. It would be intriguing if Ru$_{1-x}$Os$_x$Cl$_3$ with $x > 0.40$ features a certain aspect of the randomness-induced quantum spin liquid state.

**5. Conclusions**

We examined Os substitution effects for Ru in the Kitaev compound α-RuCl$_3$. A unique phase diagram is obtained, in which the antiferromagnetic LRO is suppressed with increasing $x$ and replaced by a spin-singlet dimer phase in a dome shape for $0.15 \leq x \leq 0.40$. Then, a magnetic SRO emerges as a ground state for $x >$ 0.40. It is suggested that Ru–Os pairs are responsible for the spin-singlet dimer formation in the presence of pseudo-threefold rotational symmetry around a substituted Os atom in a solid solution only at low substitutions. This is a rare example of dimer crystallization (MOC) induced by elemental substitution in the highly disordered system. The magnetic SRO for $x > 0.40$ may be related to a random-singlet spin liquid.

**Acknowledgments**

ZH thanks Hikaru Kawamura for helpful discussion. The XRD experiments at BL08B in Photon Factory were performed under the approval of the Photon Factory Program Advisory Committee (Proposal No. 2020G628). This work was partly supported by the Japan Society for the Promotion of Science (JSPS) KAKENHI Grant Nos. JP18H01169 and JP20H05150 (Quantum Liquid Crystals) and the Institute for Basic Science (Grant No. IBS-R009-Y3).